\documentclass{article}

\usepackage{fullpage} 

\usepackage[english]{babel} 

\usepackage[utf8]{inputenc} 

\usepackage{amsmath} 
                     
\usepackage{amssymb}
\usepackage{graphicx}
\usepackage{bbm}
\usepackage{color}
\usepackage{geometry}

\newcommand{\footnoteremember}[2]{
  \footnote{#2}
  \newcounter{#1}
  \setcounter{#1}{\value{footnote}}
}
\newcommand{\footnoterecall}[1]{
  \footnotemark[\value{#1}]
}

\author{Frédéric Abergel\footnoteremember{myfootnote}{BNP Paribas Chair of Quantitative Finance, Ecole Centrale Paris, MAS Laboratory.}\\
        \and
        Nicolas Huth\footnote{Natixis, Equity Derivatives and Arbitrage. E-mail: nicolas.huth@natixis.com. The authors would like to thank the members of Natixis' equity derivatives quantitative R\&D team for fruitful discussions.} \footnoterecall{myfootnote}
        \and
        Ioane Muni Toke\footnoterecall{myfootnote}
}

\title{Financial bubbles analysis with a cross-sectional estimator}
\date{\today}

\begin{document}
\maketitle

\begin{abstract}
We highlight a very simple statistical tool for the analysis of financial bubbles, which has already been studied in \cite{Kaizoji2006}. We provide extensive empirical tests of this statistical tool and investigate analytically its link with stocks correlation structure.
\end{abstract}

\section*{Introduction}

Forecasting the burst of financial bubbles would be incredibly useful for many players in stock exchanges, including regulators, portfolio managers and investment banks. Fundamental indicators relying on economic analysis can be monitored. But is it possible to find some statistical regularity in market crashes? Several authors, including \cite{Lillo2000,Sornette2001,Borland2009}, have already attempted to answer this question. In this paper, we focus on a very simple statistical tool, first introduced in \cite{Kaizoji2006}, and study analytically its link with stocks correlation structure. This approach is similar to the one studied in \cite{Lillo2000,Borland2009} although different through the statistical object under consideration.

\section{A spatial survival function}

In \cite{Kaizoji2006}, an unusual and interesting statistical tool is introduced in order to study market crashes. Given $N$ stocks on a market place, a reference date\footnote{supposed to be close to the onset of a financial bubble} $t_{ref}$ and the current date $t$, we set
$$
S_N(z)=\frac{1}{N}\sum_{i=1}^N \mathbbm{1}_{\left\{X_i(t_{ref},t)>z\right\}}
$$
where $X_i(t_{ref},t):=\frac{S_i(t)}{S_i(t_{ref})}$ is the performance of stock $i$ over $\left[t_{ref},t\right]$. In the following, we shall leave the time argument for notational simplicity. Intuitively, $S_N(z)$ can be seen as the proportion of stocks displaying a greater performance than $z$, i.e. the survival function of stocks on day $t$. From this point of view, it is a measure of stocks dispersion: a slow decreasing $S_N(z)$ indicates broadly distributed performances, thus reflecting an important dispersion. Since $t_{ref}$ is supposed to be as close as possible to the onset of the bubble, $X_i(t_{ref},t)$ might be of the order of monthly or yearly returns. Short notes on similar statistical objects can be found in \cite{Kaizoji2004,Dongping2008}. This is very different from looking at daily returns as in \cite{Lillo2000,Borland2009} and might be more relevant for bubble detection since it often takes long time for a bubble to build up and for bubbling stocks to disperse.\\

Statistical properties of $S_N(z)$ are interestingly robust. The main features of this quantity are:
\begin{itemize}
	\item for $z \thicksim +\infty$, $S_N(z)\thicksim z^{-\alpha}$; 
	\item the variance of the $X_i$'s increases dramatically before crashes.
\end{itemize}

\begin{figure}
\begin{center}
\begin{tabular}{ccc}
	\includegraphics[angle=0,width=0.3\textwidth]{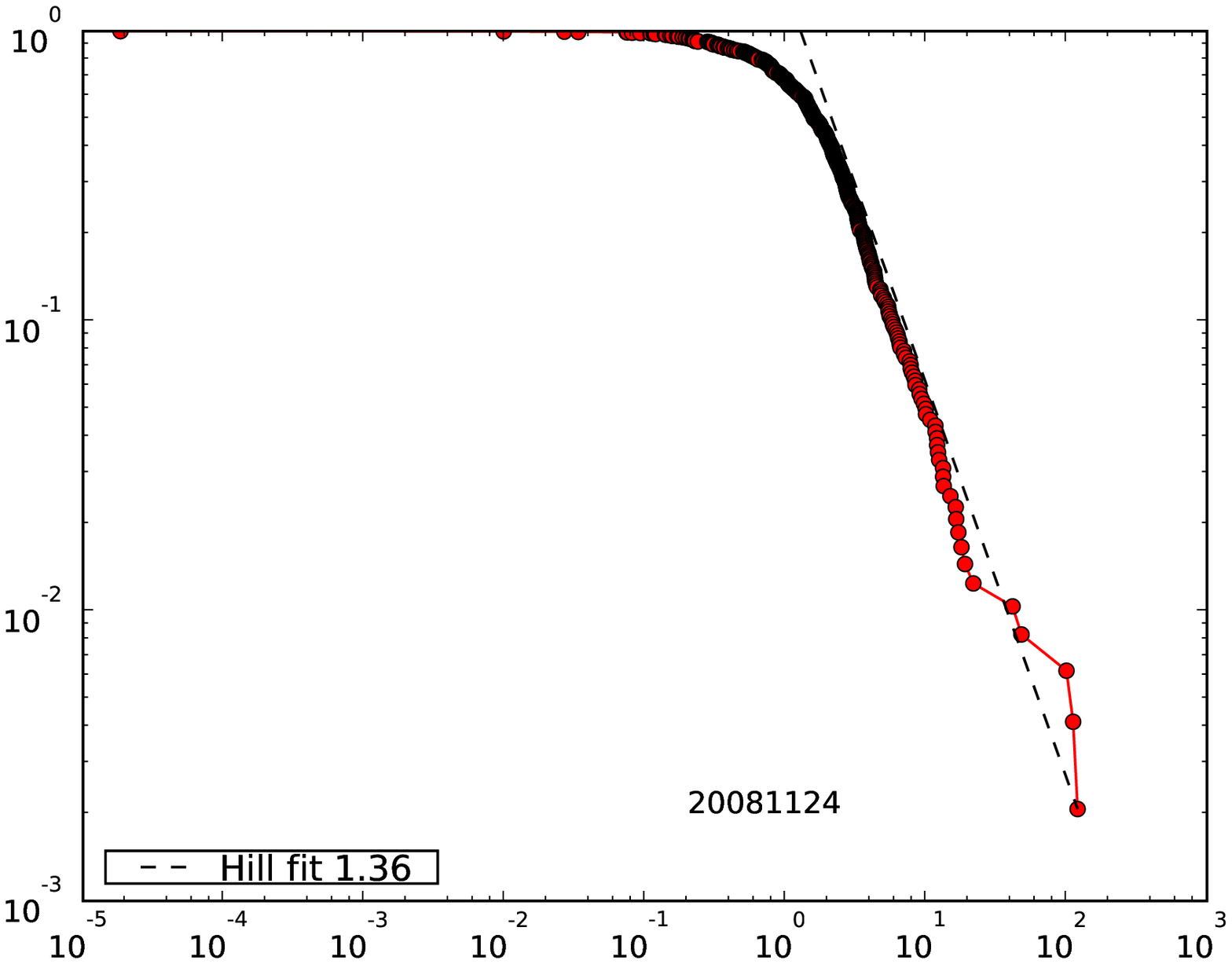}
	&
	\includegraphics[angle=0,width=0.3\textwidth]{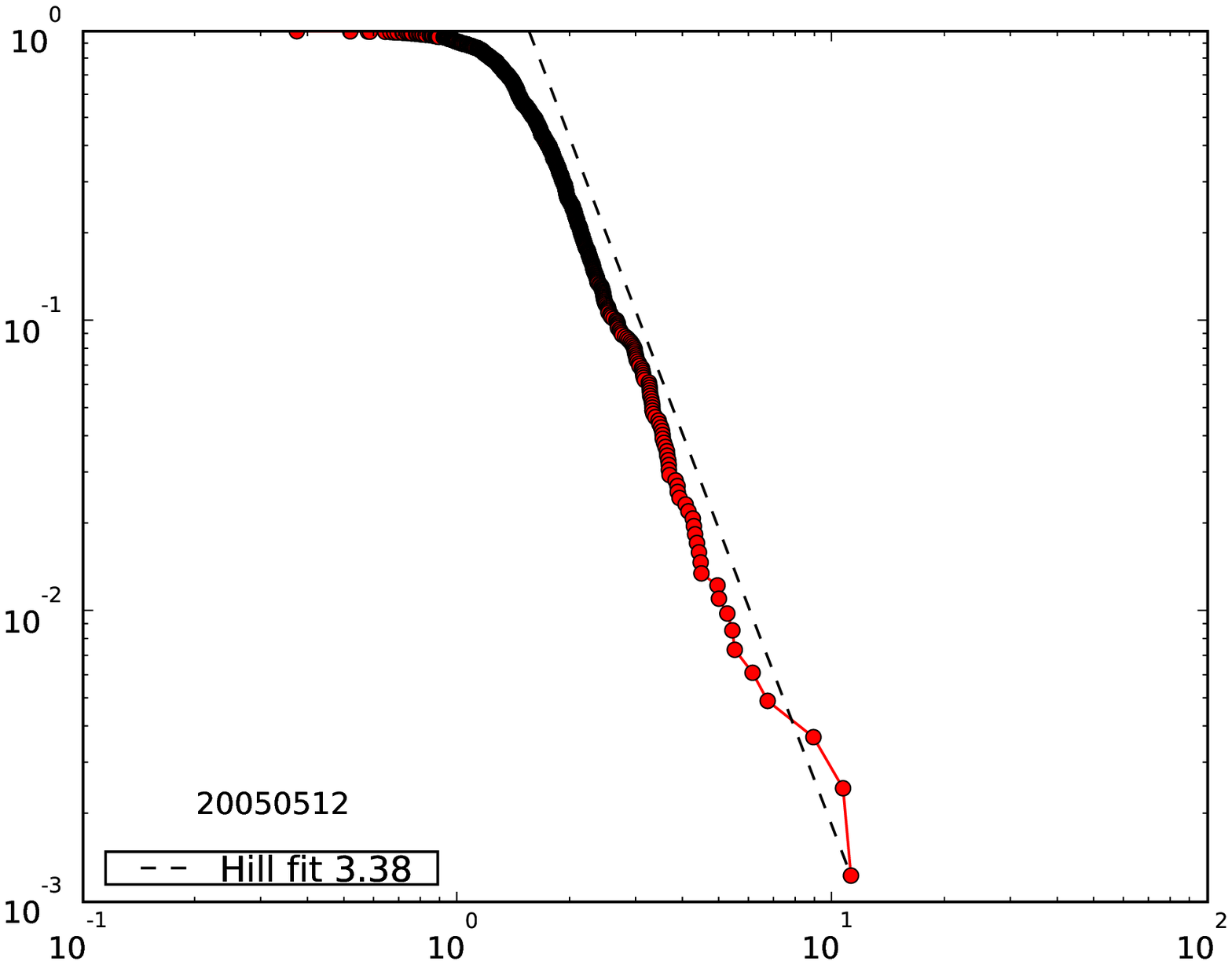}
	&
	\includegraphics[angle=0,width=0.3\textwidth]{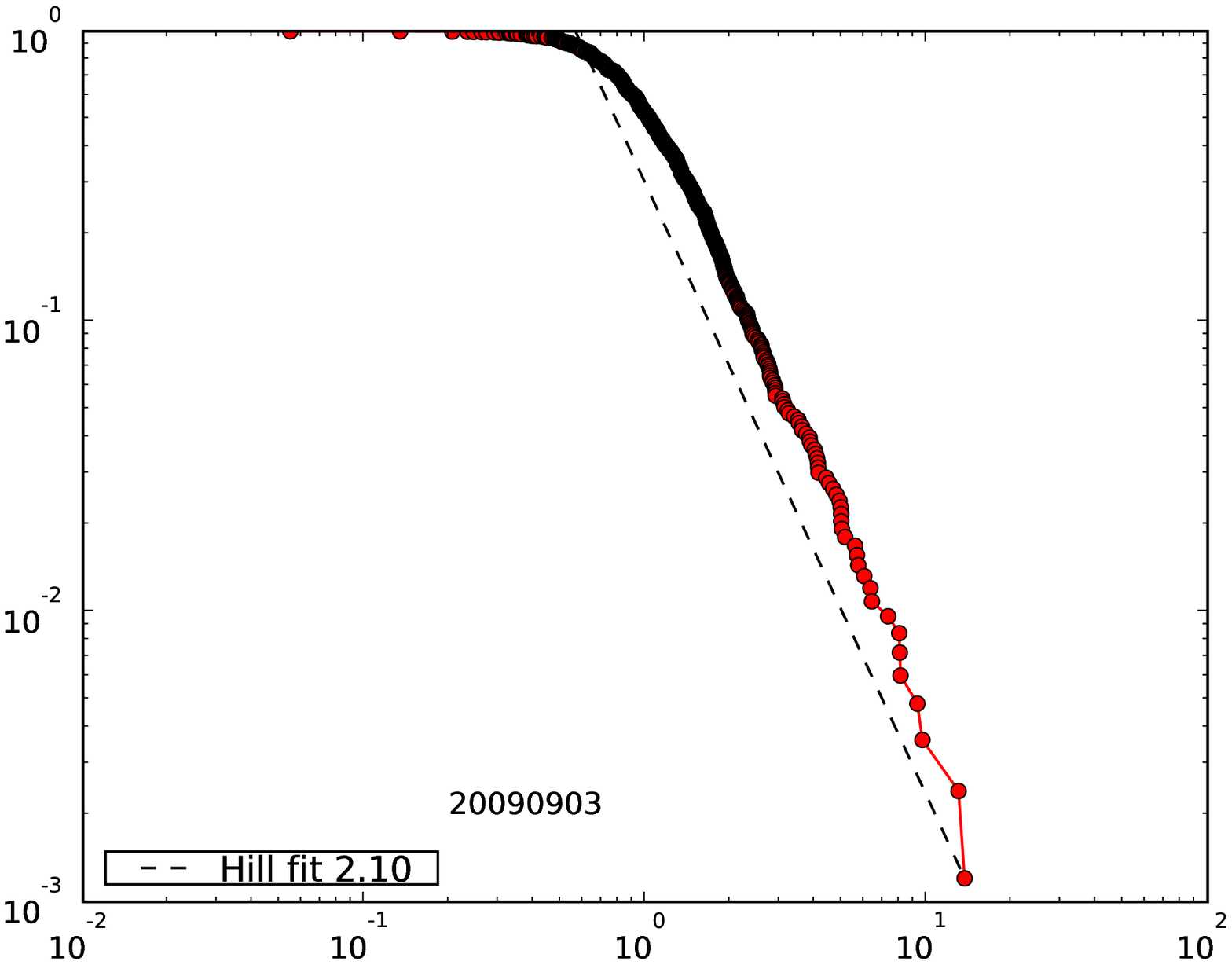}
\end{tabular}
\end{center}
\caption{Survival functions of cross-sectionnal distributions of normalized prices on three sets of stocks at random dates, in log-log scale. (left) AORD stocks on Nov 24th, 2008; (center) NYA stocks on Dec 12th, 2005; (right) SSE stocks on Sep 3rd, 2009}
\label{figure:DistributionPrices}
\end{figure}

\begin{figure}
\begin{center}
\begin{tabular}{ccc}
	\includegraphics[angle=270,width=0.3\textwidth]{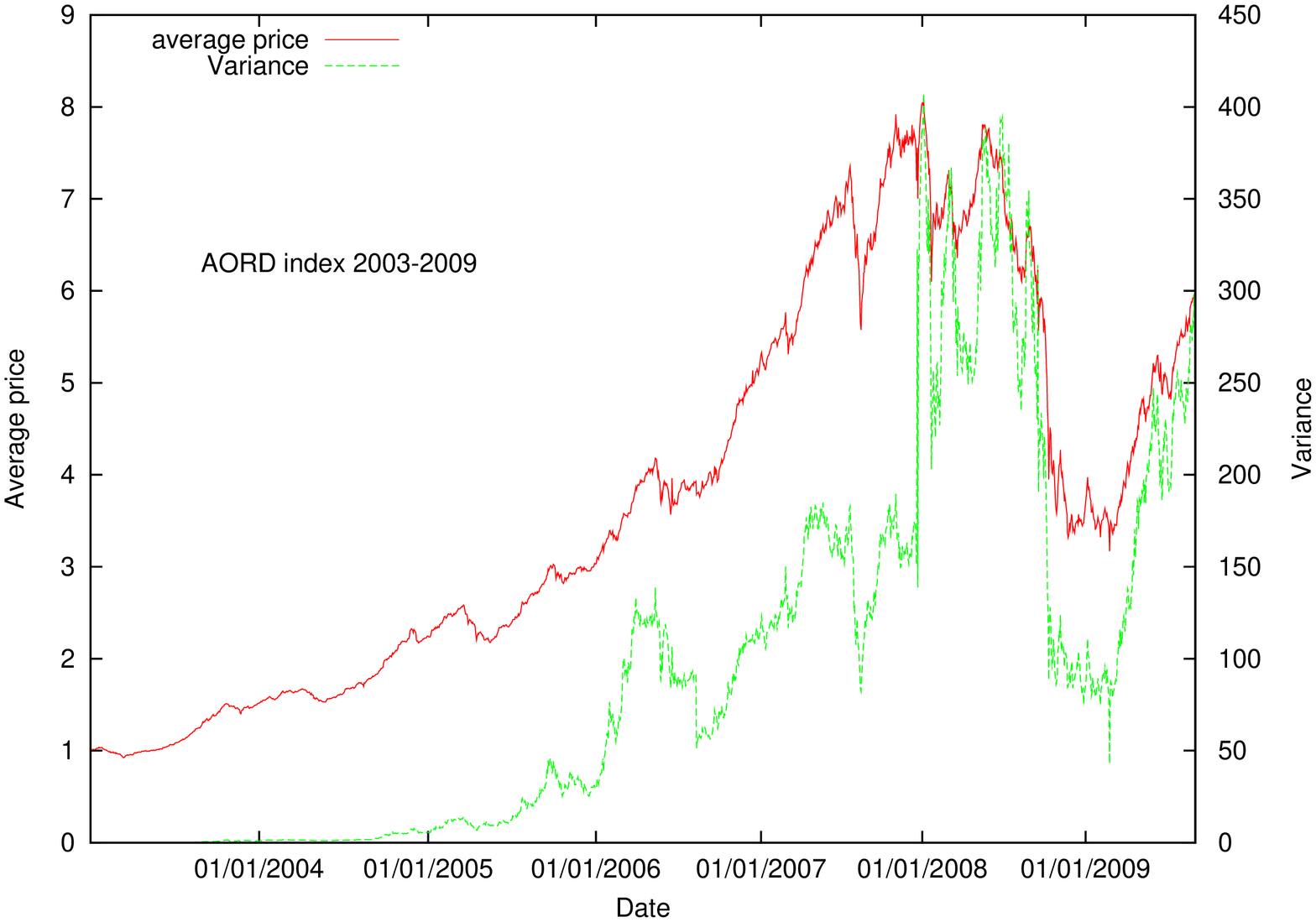}
	&
	\includegraphics[angle=270,width=0.3\textwidth]{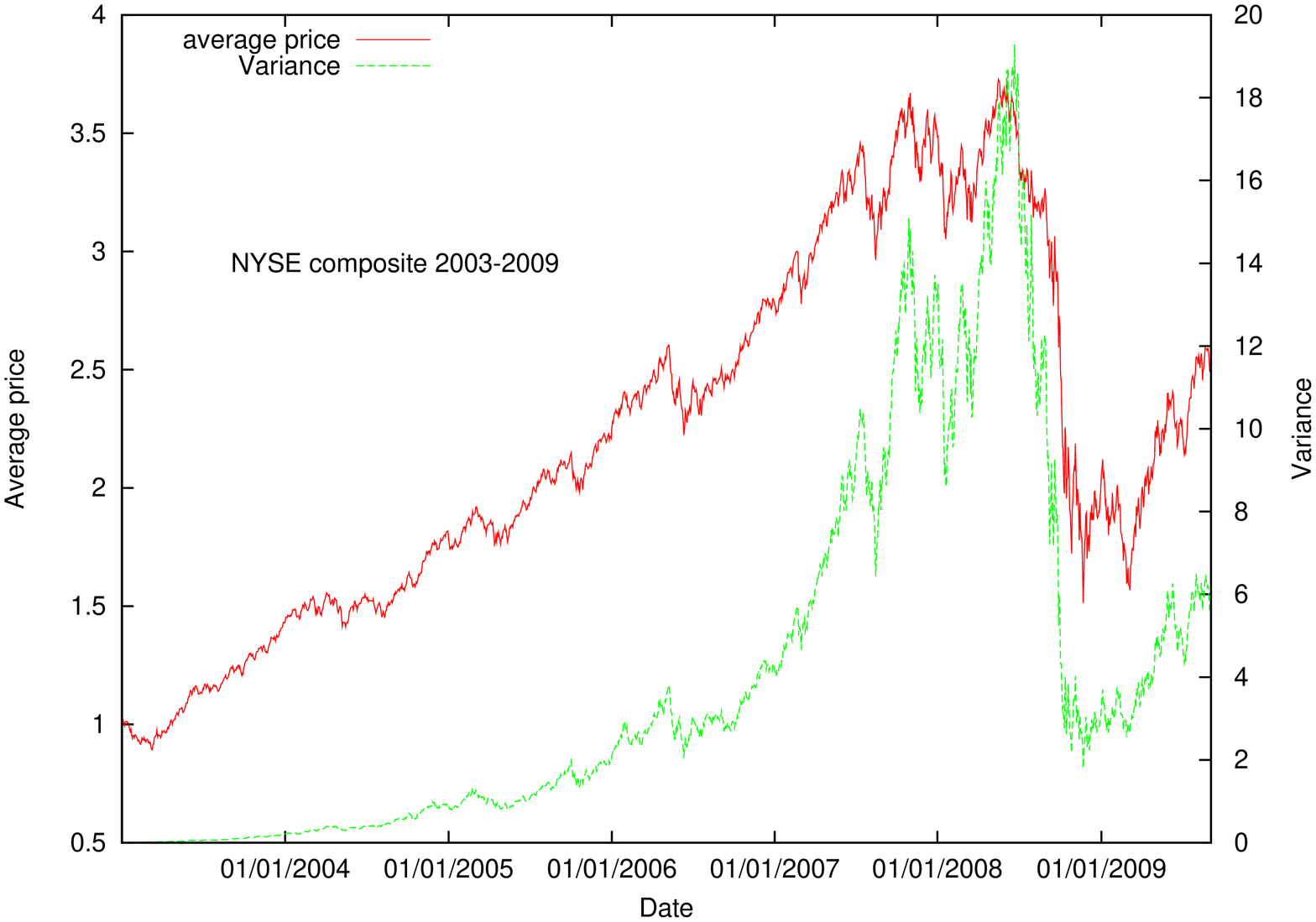}
	&
	\includegraphics[angle=270,width=0.3\textwidth]{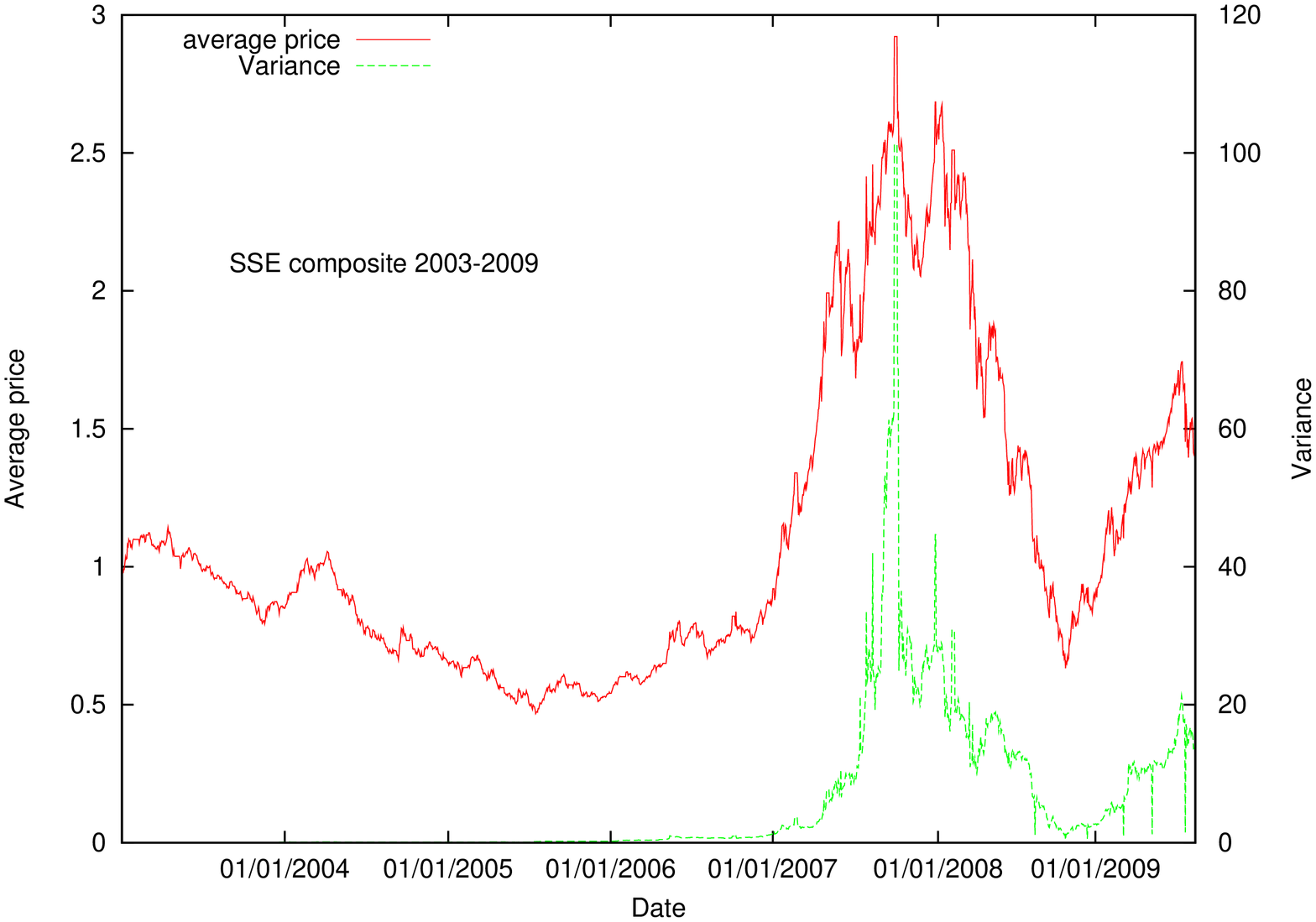}
\end{tabular}
\end{center}
\caption{Evolutions of the mean (``average price'') and variance of the $X_i$'s on three sets of stocks from Jan 1st, 2003 to Sep 3rd, 2009.(left) AORD stocks; (center) NYA stocks ; (right) SSE stocks}
\label{figure:PriceVarianceTimeseries}
\end{figure}

These two features are robust with respect to the choice of the arbitrary reference date $t_{ref}$ and are valid over several financial crashes. We illustrate these two facts on figures \ref{figure:DistributionPrices} and \ref{figure:PriceVarianceTimeseries}. We use daily close prices of three different sets of stocks: the stocks composing the Australian Stock Exchange All Ordinaries index (AORD, $500$ stocks); the stocks composing the New York Stock Exchange Composite index (NYA, $1800$ stocks); the stocks composing the Shanghai Composite index (SSE, $900$ stocks). We choose the first trading day of 2003 as our date $t_{ref}$. As for the first fact, figure \ref{figure:DistributionPrices} shows three examples of distributions of $X_i$'s at random dates. It appears that the power-law tail is indeed a good fit as the normalized prices grow.
As for the second fact, we plot on figure \ref{figure:PriceVarianceTimeseries} the timeseries of the average and variance of the $X_i$'s. It appears clearly that on all stocks, the variance grows dramatically as the bubble inflates, especially prior to crashes.

\begin{figure}
\begin{center}
\begin{tabular}{cc}
	\includegraphics[angle=0,width=0.45\textwidth]{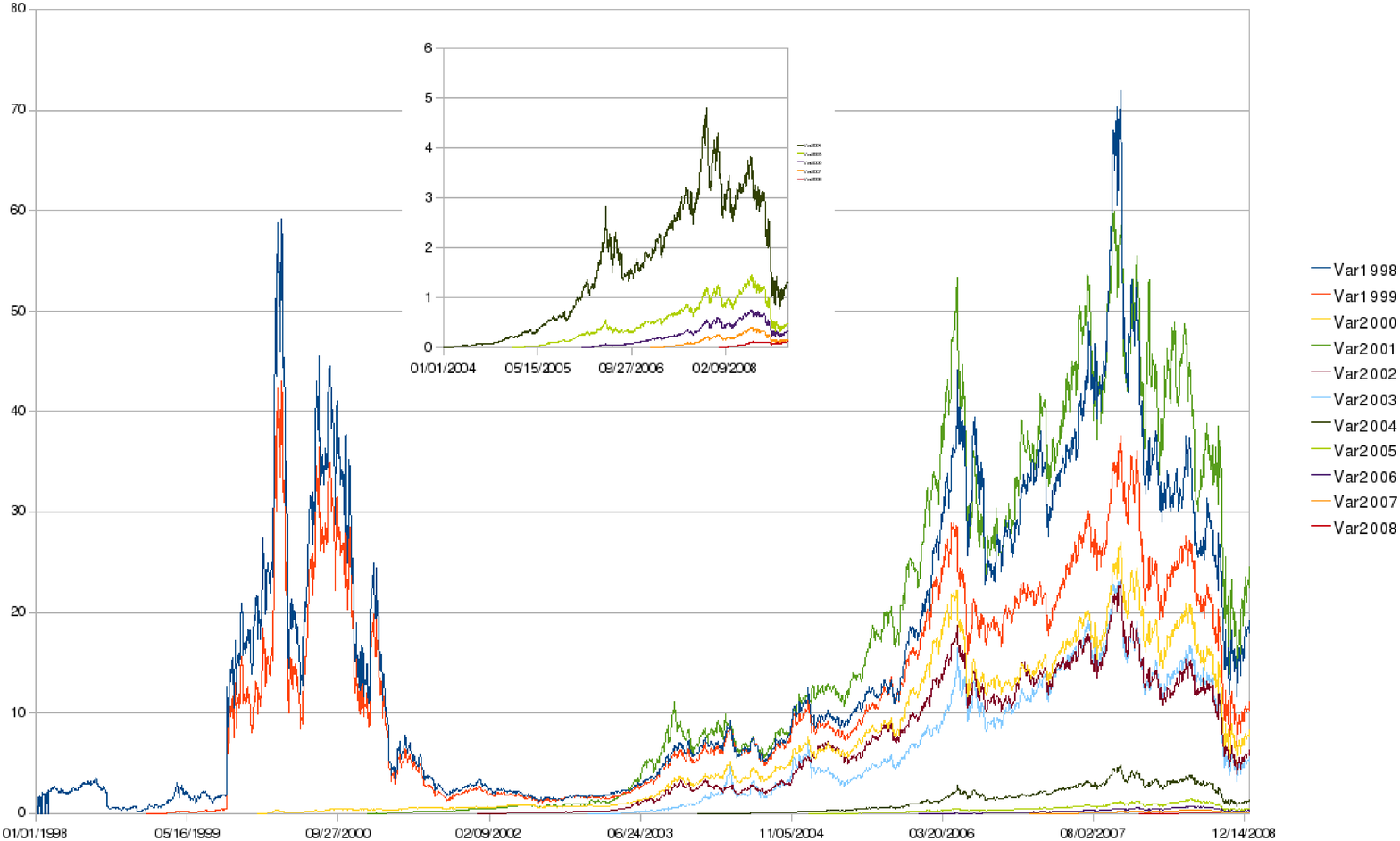}
	&
	\includegraphics[angle=0,width=0.45\textwidth]{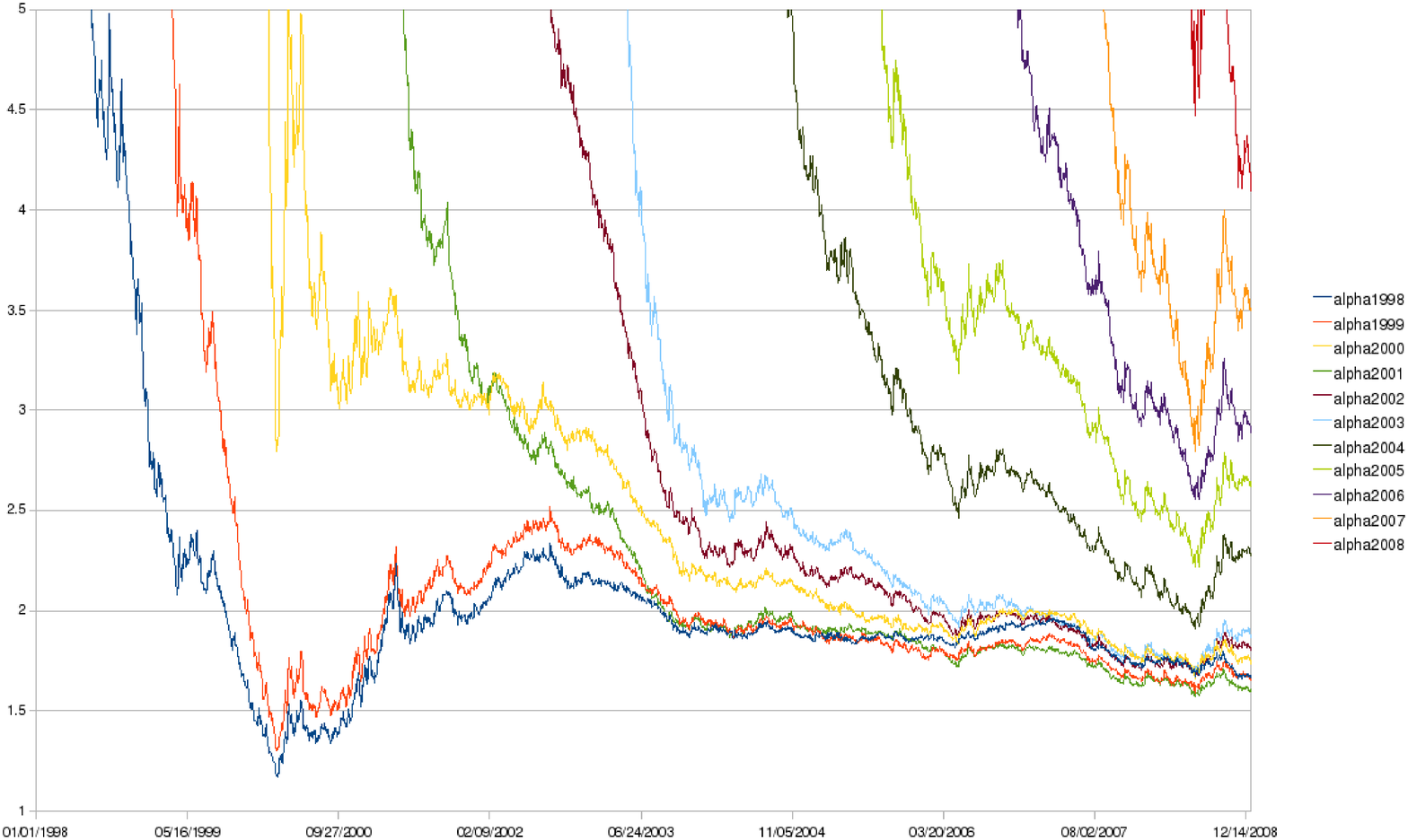}
\end{tabular}
\end{center}
\caption{(left) Variance of the normalized prices $X_i$ for eleven different reference dates $t_{ref}$. (right) Hill estimator of the Pareto exponent of the tail of distribution of the normalized prices $X_i$ for eleven different reference dates $t_{ref}$. (all) These graphs are computed using daily close prices of the stocks composing the US Russell3000 index ($3000$ stocks).}
\label{figure:PriceVarianceTrefStability}
\end{figure}

We also observe that theses properties are robust with respect to the choice of $t_{ref}$: on figure \ref{figure:PriceVarianceTrefStability} (left), timeseries of the variance of the $X_i$'s are plotted for eleven different reference dates (first trading day of the year from 1998 to 2008). It appears that the bursts of variance exists whatever the date $t_{ref}$. For the series that begin before 2000, the Internet bubble is clearly visible. The recent 2007 bubble is also visible for all series, even for the most recent reference dates (see zoom in inset).

In \cite{Kaizoji2006}, the author argues that the power-law exponent $\alpha$ dives towards $1$ before crashes, leading to the divergence of the mean performance and thus to the burst of the financial bubble. We find this threshold to be not robust with respect to $t_{ref}$: on figure \ref{figure:PriceVarianceTrefStability} (right), the Hill estimator of the Pareto exponent of the tail of the distributions of the $X_i$ is plotted, and the value of the index clearly depends on the reference date $t_{ref}$.
However, the decrease in $\alpha$ and the rise of the variance are linked since the variance of a random variable distributed as a power-law of index $\alpha$ is given by $\frac{\alpha}{(\alpha-1)^2(\alpha-2)}$. On figure \ref{figure:PriceVarianceTrefStability} (right), we observe that the Pareto exponent shows local minima and sharp peaks prior crashes. For example, first minimum is obtained on March 6th 2000 for all (started) time series, and the maximum of the Internet bubble is observed on March 9th 2000. More recently, all times series exhibit a sharp minimum on June 30th-July 1st 2008, prior to the large market dive of September 2008 (market height in 2008 is attained on June 6th).

\section{Link with the covariance structure}

On the one hand, financial bubbles happen when market quotes of certain sectors of the economy are booming, thus increasing the dispersion over whole stocks. On the other hand, it is common knowledge that market crashes are associated to bursts of volatility and correlation. How can we relate these two phenomena? Empirically, the rise in dispersion is given by the increase in the variance of performances $V_N$, which can be computed as follows
$$
V_N=\int_{0}^{+\infty}z^2 s_N(z){\rm d}z-\left(\int_{0}^{+\infty}z s_N(z){\rm d}z\right)^2
$$
where $s_N(z)=-\frac{{\rm d}}{{\rm d}z}S_N(z)=\frac{1}{N}\sum_{i=1}^N \delta(X_i-z)$ is the empirical probability density function associated to performances. It is easily seen that
$$
V_N=\frac{1}{N}\sum_{i=1}^N\left(X_i-\bar{X}_N\right)^2=\frac{1}{2N^2}\sum_{i,j=1}^N\left(X_i-X_j\right)^2
$$
with $\bar{X}_N:=\frac{1}{N}\sum_{i=1}^N X_i$. Since $V_N$ is a sum of random variables, we would like to apply a convergence theorem such as the law of large numbers. However, we must be cautious since the $X_i$'s are correlated and not identically distributed. $V_N$ will converge towards its mean if its variance is asymptotically nil. We set $m_i:=\mathbb{E} \left(X_i\right)$, $\sigma^2_i:=\mathbb{V}\text{ar} \left(X_i\right)$, $\rho_{ij}:=\mathbb{C}\text{orr} \left(X_i,X_j\right)$. Straightforward computations show that

\begin{align*}
&\mathbb{E}\left(V_N\right)=\frac{1}{N}\sum_{i=1}^N \sigma^2_i-\frac{1}{N^2}\sum_{i=1}^N \sum_{j=1}^N\rho_{ij}\sigma_i\sigma_j + \frac{1}{N}\sum_{i=1}^N m_i^2-\left(\frac{1}{N}\sum_{i=1}^N m_i\right)^2\\
&\mathbb{E}\left(V_N^2\right)=\frac{1}{N^2}\sum_{i=1}^N\mathbb{E}\left(\left(X_i-\bar{X}_N\right)^4\right)+\frac{2}{N^2}\sum_{i=1}^N\sum_{j>i}\mathbb{E}\left(\left(X_i-\bar{X}_N\right)^2\left(X_j-\bar{X}_N\right)^2\right)
\end{align*}
These two quantities exist provided that $c_{ij}^{\alpha,\beta}:=\mathbb{E}\left(\left|X_i^\alpha X_j^\beta\right|\right)<\infty$ for all $(i,j)$ and $(\alpha,\beta) \in {\cal E}:=\left\{(\gamma,\delta)\in \left\{0,1,2,3,4\right\}^2\right.$
$\left. \Big| \gamma+\delta\leq 4\right\}$. 
If the sequences $\left\{c_{ij}^{\alpha,\beta},i,j=1,\ldots,N, (\alpha,\beta) \in {\cal E}\right\}$ are such that $\lim_{N \rightarrow +\infty}\mathbb{V}\text{ar} \left(V_N\right)=0$, then
$$
\mathbb{P}-\lim_{N \rightarrow +\infty} V_N = \lim_{N \rightarrow +\infty} \left(\frac{1}{N}\sum_{i=1}^N \sigma^2_i-\frac{1}{N^2}\sum_{i=1}^N \sum_{j=1}^N\rho_{ij}\sigma_i\sigma_j + \frac{1}{N}\sum_{i=1}^N m_i^2-\left(\frac{1}{N}\sum_{i=1}^N m_i\right)^2\right)
$$
The above equation relates the dispersion of stocks $V_N$ to their variance-covariance (and mean) structure. The dispersion $V_N$:

\begin{itemize}
	\item increases as individual variances increase: if stocks are individually very unstable, then it is very likely that the whole market will be so;
	\item increases as covariances, in particular as correlations, decrease: the more anti-correlated stocks are, the more distant each pair will be, so that the whole dispersion increases.\\
\end{itemize}
Both effects are illustrated on table \ref{table:CorrelationSimulation}. We simulate a Gaussian vector with mean zero in dimension $N=1000$ a $M=100$ times, then compute the mean of $V_N$ over these $M$ simulations for different levels of correlation or standard deviation.\\

\begin{table}[!h]
\centering
\begin{tabular}{|c|c|}
\hline
$\rho$ & Mean $V_N$ \\
\hline
\hline
-1 & 1.997 \\
\hline
-0.8 & 1.808 \\
\hline
-0.6 & 1.595 \\
\hline
-0.4 & 1.393 \\
\hline
-0.2 & 1.199 \\
\hline
0 & 0.999 \\
\hline
0.2 & 0.797 \\
\hline
0.4 & 0.603 \\
\hline
0.6 & 0.400 \\
\hline
0.8 & 0.199 \\
\hline
1 & 7.430 $\times 10^{-13}$ \\
\hline
\end{tabular}
\begin{tabular}{|c|c|}
\hline
$\sigma$ & Mean $V_N$ ($\times 10^{13}$) \\
\hline
\hline
0.1 & 0.056 \\
\hline
0.3 & 0.377 \\
\hline
0.5 & 1.654 \\
\hline
0.7 & 3.052 \\
\hline
0.9 & 3.866 \\
\hline
1.1 & 5.895 \\
\hline
1.3 & 8.765 \\
\hline
1.5 & 14.249 \\
\hline
1.7 & 13.104 \\
\hline
1.9 & 22.164 \\
\hline
\end{tabular}
\caption{Simulation of a Gaussian vector with different levels of correlation and standard deviation. As correlation decreases, the dispersion is larger. Standard deviations (resp. correlations) are set to $1$ (resp. $0.5$) in the left (resp. right) table.}
\label{table:CorrelationSimulation}
\end{table}

Assume that the initial $X_i$'s are centered and normalized so that $m_i=0$ and $\sigma_i=1$. We are then able to give bounds on $\lim V_N$ depending on correlation. Setting $\rho_{ij}=1$ for all $i,j$ leads to $\mathbb{E} \left(V_N\right) = 0$ and $\rho_{ij}=-1$ for all $i,j$ leads to $\mathbb{E} \left(V_N\right) = 2$, as shown in table \ref{table:CorrelationSimulation}.

\section{Conclusion and further research}

We have studied empirically and analytically an indicator of bubbles build up and burst on financial markets. This statistical tool is the variance of fixed starting date performances over the stocks universe. It is quite robust with respect to the choice of the market place and the starting date. The .com and subprime bubbles are well identified by this methodology. Fundamentally, we establish a link between the building up of a bubble and anti-correlation between stocks as well as individual variances bursts.

Regarding further research, we have two things in mind:
\begin{itemize}
	\item test empirically the link we establish between bubbles and stocks variance-covariance structure;
	\item suggest an agent-based model to explain the fundamental microscopic mechanisms underlying this link between bubbles and stocks variance-covariance structure.
\end{itemize}

The first direction requires a statistical measure for the variance and correlation of $X_i(t_{ref},t)$ over $i$ at each time $t$. This problem is quite complex in practice since, unless one goes $30$ years back, one does not have historical daily data to compute them. A solution might be found in the use of high frequency data by computing the variances and correlations needed over the returns of the previous day. Furthermore, having access to these statistical quantities would be useful for normalizing returns in order to bound the market variance $V_N$, thus making it an indicator with fundamental thresholds.

Some exchange models for wealth distribution, such as \cite{Bouchaud2000}, have striking similarities with our approach and could be therefore used to understand the power law distribution of stocks ensemble and global variance burst during bubbles from a microscopic point of view towards interactions between individual stocks.

\bibliographystyle{naturemag}
\bibliography{EconophysicsColloquium2009}

\begin{thebibliography}{1}
\expandafter\ifx\csname url\endcsname\relax
  \def\url#1{\texttt{#1}}\fi
\expandafter\ifx\csname urlprefix\endcsname\relax\def\urlprefix{URL }\fi
\providecommand{\bibinfo}[2]{#2}
\providecommand{\eprint}[2][]{\url{#2}}

\bibitem{Kaizoji2006}
\bibinfo{author}{Kaizoji, T.}
\newblock \bibinfo{title}{A precursor of market crashes: Empirical laws of
  japan's internet bubble}.
\newblock \emph{\bibinfo{journal}{The European Physical Journal B}}
  \textbf{\bibinfo{volume}{50}}, \bibinfo{pages}{5 pages}
  (\bibinfo{year}{2006}).

\bibitem{Lillo2000}
\bibinfo{author}{Lillo, F.} \& \bibinfo{author}{Mantegna, R.~N.}
\newblock \bibinfo{title}{Variety and volatility in financial markets}.
\newblock \emph{\bibinfo{journal}{Physical Review E}}
  \textbf{\bibinfo{volume}{62}}, \bibinfo{pages}{6126} (\bibinfo{year}{2000}).

\bibitem{Sornette2001}
\bibinfo{author}{Sornette, D.} \& \bibinfo{author}{Johansen, A.}
\newblock \bibinfo{title}{Significance of log-periodic precursors to financial
  crashes}.
\newblock \emph{\bibinfo{journal}{Quantitative Finance}}
  \textbf{\bibinfo{volume}{1}}, \bibinfo{pages}{452–471}
  (\bibinfo{year}{2001}).

\bibitem{Borland2009}
\bibinfo{author}{Borland, L.}
\newblock \bibinfo{title}{Statistical signatures in times of panic: Markets as
  a {Self-Organizing} system}.
\newblock \emph{\bibinfo{journal}{Arxiv preprint {arXiv:0908.0111}}}
  (\bibinfo{year}{2009}).

\bibitem{Kaizoji2004}
\bibinfo{author}{Kaizoji, T.} \& \bibinfo{author}{Kaizoji, M.}
\newblock \bibinfo{title}{Power law for ensembles of stock prices}.
\newblock \emph{\bibinfo{journal}{Physica A: Statistical Mechanics and its
  Applications}} \textbf{\bibinfo{volume}{344}}, \bibinfo{pages}{240--243}
  (\bibinfo{year}{2004}).

\bibitem{Dongping2008}
\bibinfo{author}{Men, D.}, \bibinfo{author}{Wang, J.} \& \bibinfo{author}{Shao,
  J.}
\newblock \bibinfo{title}{The statistical analysis of stock prices and trading
  volumes for the chinese stock markets}.
\newblock In \emph{\bibinfo{booktitle}{Computing, Communication, Control, and
  Management, 2008. {CCCM} '08.}}, vol.~\bibinfo{volume}{3},
  \bibinfo{pages}{37--41} (\bibinfo{year}{2008}).

\bibitem{Bouchaud2000}
\bibinfo{author}{Bouchaud, J.~P.} \& \bibinfo{author}{Mezard, M.}
\newblock \bibinfo{title}{Wealth condensation in a simple model of economy}.
\newblock \emph{\bibinfo{journal}{Physica A: Statistical Mechanics and its
  Applications}} \textbf{\bibinfo{volume}{282}}, \bibinfo{pages}{536–545}
  (\bibinfo{year}{2000}).

\end{thebibliography}

\end{document}